\documentclass[journal=acsodf,manuscript=article]{achemso}

\usepackage[T1]{fontenc}
\usepackage[utf8]{inputenc}
\usepackage{graphicx}
\usepackage{amsmath,amssymb}
\usepackage{booktabs}

\author{Marcos V. N. da Costa}
\affiliation[PPGCIMA]{Postgraduate Program in Materials Science (PPGCIMA), Faculty UnB Planaltina (FUP), University of Bras\'ilia, 73345-010 Bras\'ilia, DF, Brazil}
\alsoaffiliation[NanoEng]{NanoEngineering Laboratory (NanoEng), College of Technology, University of Bras\'ilia, 70910-900 Bras\'ilia, DF, Brazil}

\author{Jos\'e R. da M. Lima Filho}
\affiliation[PPGCIMA]{Postgraduate Program in Materials Science (PPGCIMA), Faculty UnB Planaltina (FUP), University of Bras\'ilia, 73345-010 Bras\'ilia, DF, Brazil}
\alsoaffiliation[NanoEng]{NanoEngineering Laboratory (NanoEng), College of Technology, University of Bras\'ilia, 70910-900 Bras\'ilia, DF, Brazil}

\author{K\'adila R. de S. Oliveira}
\affiliation[PPGCIMA]{Postgraduate Program in Materials Science (PPGCIMA), Faculty UnB Planaltina (FUP), University of Bras\'ilia, 73345-010 Bras\'ilia, DF, Brazil}
\alsoaffiliation[NanoEng]{NanoEngineering Laboratory (NanoEng), College of Technology, University of Bras\'ilia, 70910-900 Bras\'ilia, DF, Brazil}

\author{Raphael M. Tromer}
\affiliation[IF]{Institute of Physics, University of Bras\'ilia, 70910-900 Bras\'ilia, DF, Brazil}
\alsoaffiliation[NanoEng]{NanoEngineering Laboratory (NanoEng), College of Technology, University of Bras\'ilia, 70910-900 Bras\'ilia, DF, Brazil}

\author{Marcelo L. Pereira Junior}
\email{marcelo.lopes@unb.br}
\affiliation[ENE]{Department of Electrical Engineering, College of Technology, University of Bras\'ilia, 70910-900 Bras\'ilia, DF, Brazil}
\alsoaffiliation[PPGCIMA]{Postgraduate Program in Materials Science (PPGCIMA), Faculty UnB Planaltina (FUP), University of Bras\'ilia, 73345-010 Bras\'ilia, DF, Brazil}
\alsoaffiliation[NanoEng]{NanoEngineering Laboratory (NanoEng), College of Technology, University of Bras\'ilia, 70910-900 Bras\'ilia, DF, Brazil}

\title{Machine-Learning Exploration of Defect Topologies and Thermodynamic Stability in Graphene with Atomic Vacancies}


\begin{document}

\begin{tocentry}
\includegraphics[width=\linewidth]{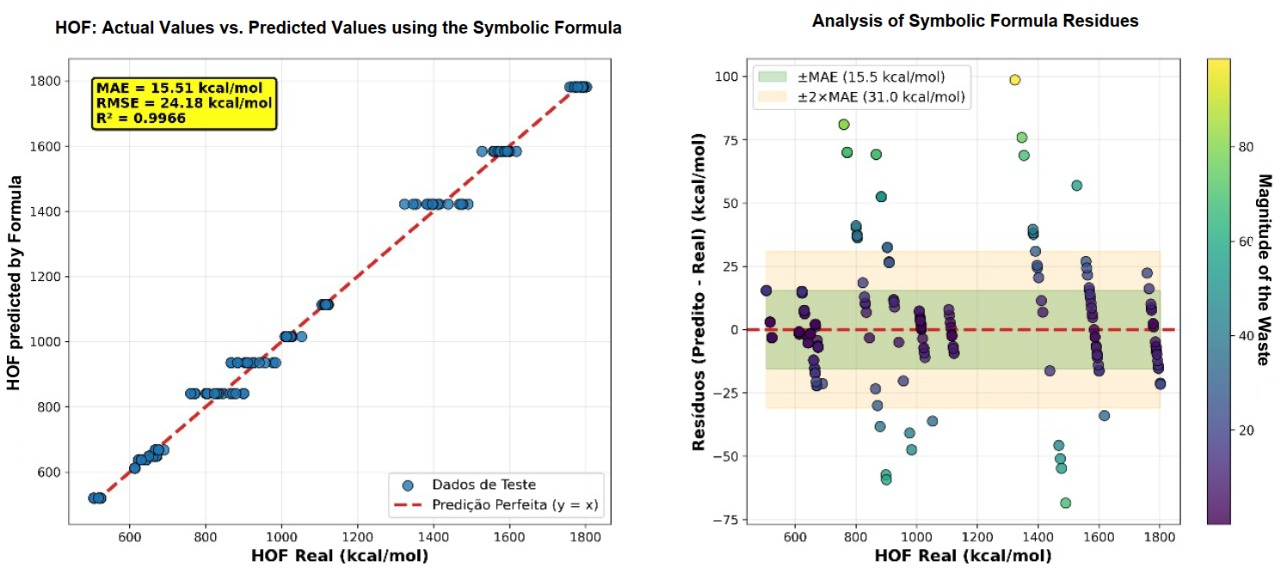}
\end{tocentry}

\begin{abstract}
Atomic vacancies and vacancy aggregates control the thermodynamic stability and the functional response of graphene, yet the configurational space spanned by many vacancies at variable concentration and separation is too large to be mapped exhaustively by first-principles methods. Here, we map and rationalize this stability landscape by combining semiempirical atomistic thermodynamics, interpretable machine learning, and symbolic regression. Several hundred defective supercells, built from a 72-atom cell by varying the vacancy concentration and the inter-vacancy distance up to the fourth neighbor, were relaxed with the PM7 Hamiltonian in MOPAC, and the heat of formation was adopted as the stability metric. Each structure was encoded with the Dynamic Collision Fingerprint, a translationally and rotationally invariant descriptor that maps the local topology onto transport-like statistics of virtual probe particles. A gradient boosted decision tree model, optimized by Bayesian hyperparameter search, reproduces the heat of formation of an independent test set with a root mean squared error of approximately 23.5~kcal/mol and no evidence of overfitting, and a SHAP analysis identifies the defect concentration and the inter-vacancy distance as the two variables that dominate the stability. Symbolic regression then condenses the learned mapping into a compact closed-form expression that reproduces the heat of formation with a coefficient of determination of $R^{2} = 0.9966$, a mean absolute error of 15.51~kcal/mol, and a root mean squared error of 24.18~kcal/mol. The workflow turns a high-dimensional structure--stability problem into an interpretable analytical law, providing a transferable route to rational defect engineering in two-dimensional materials.
\end{abstract}

\section{Introduction}

Carbon-based materials occupy a central position in materials science, and graphene in particular combines a high carrier mobility, a large specific surface area, and a remarkable mechanical strength that have motivated its integration into electronic, sensing, and energy-storage devices \cite{vieirasegundo2016grafeno, banhart2011structural}. Real graphene samples are never defect free, and the deliberate introduction of structural imperfections has evolved from an unavoidable byproduct of growth into a strategy for tuning the electronic, magnetic, mechanical, and chemical response of the material \cite{trevethan2014vacancy, xiong2021structural, lin2015defect}. Atomic vacancies are the prototypical example, since removing carbon atoms leaves under-coordinated sites that carry local magnetic moments \cite{Yazyev2007}, reshape the electronic structure, and soften the lattice, while irradiation provides a controllable route to introduce them at a chosen density \cite{Krasheninnikov2010}. The deliberate use of such imperfections to program material properties is the basis of defect engineering in two-dimensional systems \cite{lin2015defect, bhatt2022various}.

The thermodynamic stability of a defective configuration is conveniently quantified by its heat of formation, which condenses the energetic cost of bond breaking and lattice relaxation into a single scalar \cite{banhart2011structural}. A reconstructed monovacancy in graphene carries a formation energy of the order of 7.5~eV, set by the saturation of two of the three dangling bonds left by the missing atom \cite{ElBarbary2003}, a value that even semilocal density functional theory underestimates relative to quantum Monte Carlo benchmarks \cite{Thomas2022}. Crucially, the heat of formation of a multivacancy system is not the sum of isolated-defect contributions. When two monovacancies approach, they coalesce into a divacancy that eliminates the dangling bonds and reconstructs into pentagon and heptagon rings, so that the divacancy is markedly more stable than two separated monovacancies \cite{Lee2005, banhart2011structural}. The associated strain fields are long ranged and bias the migration and the relative arrangement of neighboring defects \cite{trevethan2014vacancy}, an interplay also documented for the interaction and ordering of point defects in other crystals \cite{vandermeer2017diffusion, landesman1985ordering, Pietrucci2008, Bogicevic2003}. Most studies of graphene nevertheless address isolated vacancies or single pairs in specific arrangements \cite{tournus2005vacancy, gao2019thermodynamic, li2021vacancy}, and a systematic account of how vacancy concentration and inter-vacancy distance jointly control the heat of formation is still missing. The central physical question, namely how the cooperative and non-additive coupling between defects shapes the stability landscape, therefore remains open.

Answering this question requires sampling a configurational space that grows rapidly with the number and the arrangement of vacancies, which is precisely where first-principles methods become impractical. Density functional theory provides the reference accuracy for defective carbon, but its steep scaling with system size and the need to fully relax each defective supercell make a broad survey across vacancy sizes and separations expensive \cite{Zhou2013, Zhou2017}. Semiempirical quantum-chemical methods offer a strategic intermediate route, since the neglect of the heavier electron-repulsion integrals within the NDDO formalism, combined with parameters fitted to experimental and high-level data, reduces the cost by orders of magnitude while retaining a consistent description of the heat of formation \cite{stewart2007optimization, stewart2013optimization, cui2014density}. The reliability of semiempirical and tight-binding schemes for the global exploration of defective and cluster structures has been examined in several contexts \cite{galvao2023confiabilidade, rapacioli2011extensions, grimme2019exploration, pracht2020automated}, and the PM7 Hamiltonian in particular enables the geometry relaxation and the thermodynamic characterization of the several hundred defective supercells required to populate the configurational space studied here.

Generating a large dataset displaces the difficulty toward its analysis, because the relationship between defect geometry and stability is non-linear and high-dimensional and is not accessible to direct inspection. Supervised machine learning addresses this regime when the atomic structure is encoded into a fixed-length, symmetry-invariant representation \cite{butler2018machine, schmidt2019recent, ong2021accelerating}, an approach that has reproduced the energetics of carbon at near-ab-initio accuracy \cite{Deringer2017} and has been used to explore the configurational space of defective lattices \cite{birschitzky2022machine}. Descriptor design is the decisive step, and established representations such as atom-centered symmetry functions and the smooth overlap of atomic positions encode the local environment in a translationally and rotationally invariant form \cite{Behler2007, Bartok2013}. The Dynamic Collision Fingerprint recently introduced for two-dimensional materials follows the same principle but encodes the local topology through the collision statistics of virtual probe particles, mapping the geometric environment of the vacancies onto transport-like quantities that carry a transparent physical meaning \cite{Tromer2025, Tromer2026, Choudhary2018, Dau2023, NaTalang2026}. High-performance models such as gradient boosted decision trees then learn the structure--stability mapping with high fidelity, but their ensemble nature obscures the mechanistic content of the prediction. This opacity is resolved on two complementary levels, first by attributing the prediction to individual descriptors through SHAP values rooted in cooperative game theory \cite{lundberg2017unified, lundberg2020local}, and second by distilling the learned mapping into an explicit analytical law through symbolic regression \cite{schmidt2009distilling, cranmer2020discovering, brunton2016discovering, udrescu2020ai, guimera2020scientific}. Together, these tools convert a black-box predictor into a compact and physically interpretable description of defect stability.

In this work, we address the open question of how concentration and separation jointly govern the stability of vacancies in graphene, and we do so with a workflow that links semiempirical thermodynamics to an interpretable analytical law. We generate several hundred defective supercells with systematically varied vacancy concentration and inter-vacancy distance, compute their heat of formation with the PM7 Hamiltonian in MOPAC, and encode every structure with the Dynamic Collision Fingerprint. A gradient boosted decision tree model optimized by Bayesian hyperparameter search predicts the heat of formation with high accuracy, a SHAP analysis identifies the defect concentration and the inter-vacancy distance as the dominant variables, and symbolic regression condenses the mapping into a closed-form expression that reproduces the heat of formation with $R^{2} = 0.9966$. The representative configurations and the resulting distribution of the heat of formation that define the dataset are summarized in Figure~\ref{fig:vacancies_hof}.

\section{Methodology}

The defective structures were built from a square graphene supercell containing 72 carbon atoms, a size chosen to minimize edge effects while keeping the semiempirical relaxation of several hundred configurations tractable. Vacancies were introduced by removing carbon atoms from the pristine lattice, and the dataset was organized around two control variables. The first is the defect concentration, taken as the number of vacancies per supercell and varied over the representative values of 2, 4, and 7. The second is the inter-vacancy distance, taken as the separation between the vacancies and varied from first to fourth neighbors, so that both the short-range overlap of defect cores and the more subtle long-range elastic coupling were sampled. This construction yields several hundred distinct defect topologies spanning the relevant range of concentration and separation.

Each configuration was relaxed and characterized with semiempirical calculations performed in the MOPAC package using the PM7 Hamiltonian, which is based on the NDDO approximation with parameters fitted to experimental and density functional data and is well suited to the description of carbon-based systems \cite{stewart2013optimization}. The geometries were fully optimized, and a complete thermodynamic analysis was carried out to obtain the heat of formation together with the enthalpy, the entropy, the heat capacity, and the vibrational frequencies. The heat of formation was adopted as the target property for the machine-learning models because it provides a direct measure of the thermodynamic stability of the defective supercell.

The relaxed structures were encoded with the Dynamic Collision Fingerprint (DCF) descriptor using the parameter set proposed in the original reference \cite{Tromer2025}. The DCF simulates the trajectories of virtual probe particles that collide with the atomic framework and converts the resulting collision statistics into a fixed-length feature vector that is invariant under translation and rotation. The descriptor maps the local geometry of the vacancies onto transport-like quantities, including an effective diffusivity, a mean relaxation time, a mean free path, and angular-disorder measures, which carry a transparent interpretation of the openness and the connectivity of the defective network \cite{Tromer2025, Tromer2026}. The final dataset combines the DCF feature vector of every structure with its heat of formation and the two control variables.

The non-linear mapping between the descriptors and the heat of formation was learned with the Extreme Gradient Boosting (XGBoost) algorithm, an ensemble of decision trees in which each new tree minimizes the residual error of the previous ones \cite{chen2016xgboost}. The model is well suited to heterogeneous structure--property data and provides effective control of overfitting while capturing the intricate non-linear dependence of stability on defect geometry. The hyperparameters, including the maximum tree depth, the learning rate, the number of estimators, the subsampling and column-sampling ratios, and the L1 and L2 regularization terms, were optimized with the Optuna framework, which implements a sequential Bayesian search based on Tree-structured Parzen Estimators and concentrates the sampling on the most promising regions of the hyperparameter space \cite{akiba2019optuna}. The dataset was randomly split into 80\% for training and 20\% for testing, so that the predictive accuracy was always assessed on configurations not seen during training.

The interpretation of the optimized model was carried out at two levels. The relative contribution of each variable to the prediction was quantified with SHAP values, which assign a marginal contribution derived from cooperative game theory and thereby expose the global importance ranking, the local effect of each variable, and the non-linear interactions among them \cite{lundberg2017unified, lundberg2020local}. Building on the hierarchy revealed by SHAP, symbolic regression based on genetic programming was then applied to the most relevant variables, searching the space of elementary algebraic and transcendental operations under a parsimony criterion in order to distill the learned mapping into a compact closed-form expression for the heat of formation \cite{schmidt2009distilling, udrescu2020ai}.

\section{Results and Discussion}

\begin{figure}[htbp]
\centering
\includegraphics[width=\linewidth]{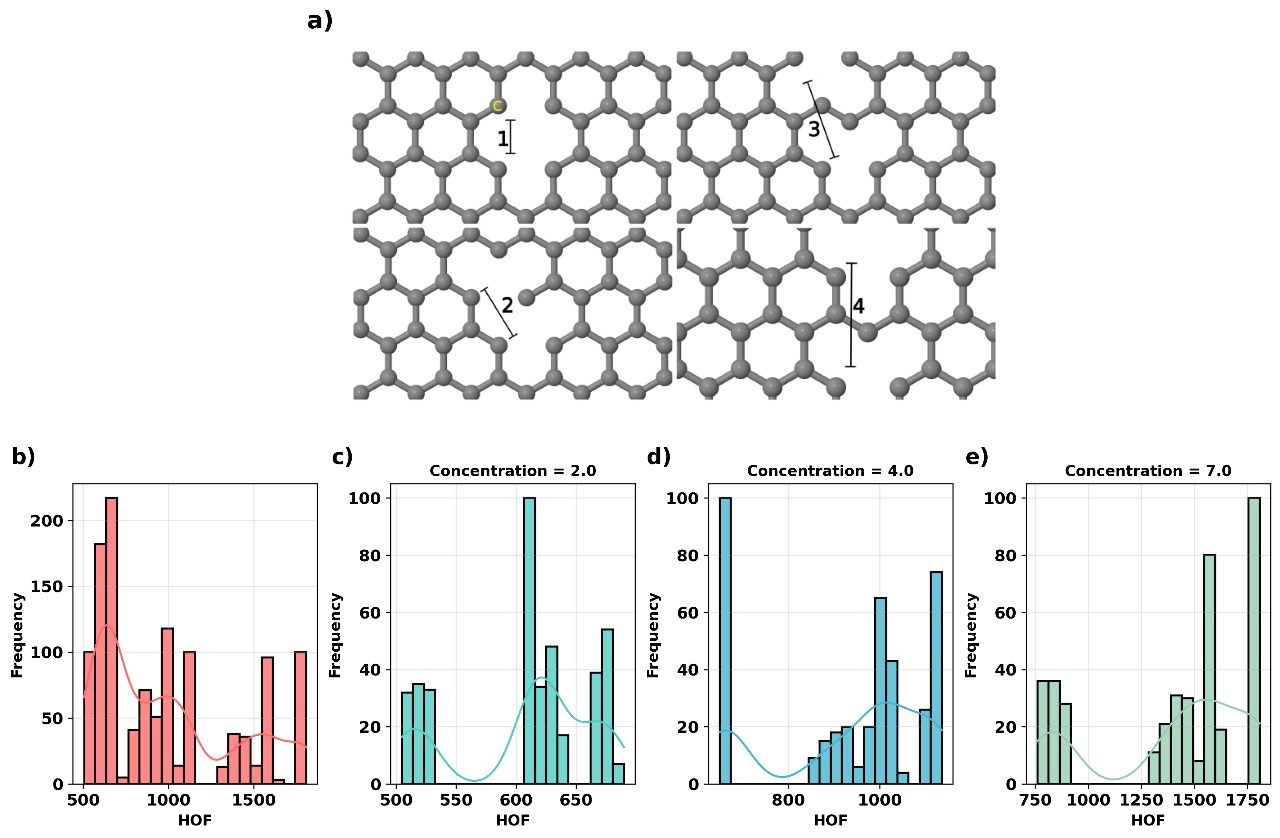}
\caption{Dataset construction. (a) Representative defective graphene configurations, with a minimal vacancy in panel (1) and vacancies at increasing separation in panels (2) to (4). (b) Global distribution of the heat of formation for all configurations. (c, d, e) Distribution of the heat of formation resolved by defect concentration, in order of increasing concentration.}
\label{fig:vacancies_hof}
\end{figure}

The construction of the dataset rests on a controlled variation of vacancy concentration and inter-vacancy distance, and the resulting spread of stabilities provides the first view of the structure--stability landscape that the subsequent analysis aims to rationalize. Figure~\ref{fig:vacancies_hof} presents representative defective configurations together with the distribution of the heat of formation resolved by defect concentration.

Figure~\ref{fig:vacancies_hof}(a) illustrates the generation protocol. Panel (1) shows a minimal vacancy created by removing one carbon pair from the pristine 72-atom supercell, while panels (2) to (4) show configurations in which vacancies of the same type are placed at increasing separation, from nearest neighbors in (2) to the largest distance considered in (4). For every configuration of this type, the PM7 relaxation in MOPAC yields the complete thermodynamic description of the system, and the relaxed atomic structure is encoded with the DCF descriptor. The global distribution of the heat of formation in Figure~\ref{fig:vacancies_hof}(b) is broad and multimodal, reflecting the coexistence of several concentrations and separations in the dataset. When the data are resolved by concentration, a systematic trend emerges. The lowest concentration in Figure~\ref{fig:vacancies_hof}(c) clusters in a narrow low-energy range, indicating that weakly defective configurations are, on average, more stable. As the concentration increases in Figures~\ref{fig:vacancies_hof}(d) and~\ref{fig:vacancies_hof}(e), the distribution shifts toward higher heats of formation and develops a more pronounced structure, consistent with the additional energetic penalty of removing more atoms and with the richer variety of local reconstructions that become accessible in the highly defective regime. The wide and structured spread of the heat of formation confirms the complexity of the underlying landscape and motivates the descriptor-based analysis that follows.

\begin{figure}[htbp]
\centering
\includegraphics[width=\linewidth]{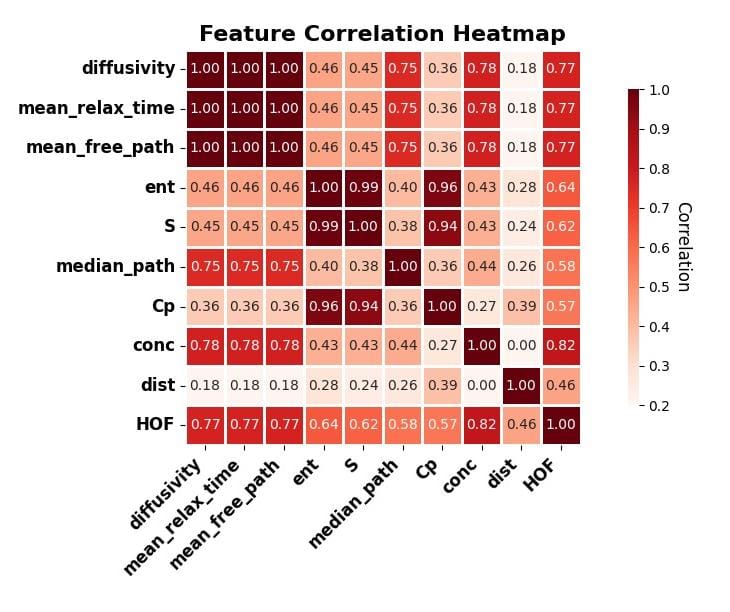}
\caption{Pearson correlation matrix between selected DCF descriptors, thermodynamic quantities, defect concentration, inter-vacancy distance, and the heat of formation. Strong correlations appear among the dynamical descriptors and between concentration and the heat of formation, while distance couples only weakly in linear terms.}
\label{fig:corr_heatmap}
\end{figure}

Before training a non-linear model, it is instructive to examine the linear structure of the dataset, since the pattern of mutual correlations among the descriptors and the target both guides the modeling strategy and exposes the redundancy inherent in the feature space. Figure~\ref{fig:corr_heatmap} reports the Pearson correlation matrix between selected DCF descriptors, the thermodynamic quantities, the defect concentration, the inter-vacancy distance, and the heat of formation.

The most prominent feature of the matrix is the nearly perfect correlation ($r \approx 1.0$) among the dynamical descriptors of the DCF, namely the effective diffusivity, the mean relaxation time, and the mean free path. This redundancy is expected from the formal definitions of these quantities, which are related to one another up to multiplicative or additive constants and therefore encode the same underlying dynamical information. Their positive correlation with the heat of formation ($r \approx 0.77$) indicates that configurations with a larger effective diffusivity, and equivalently longer mean free paths and relaxation times, tend to exhibit higher heats of formation and are thus less stable, consistent with the picture that more open or strongly perturbed networks allow easier probe motion at the cost of an increased energetic penalty. A second correlated block is formed by the purely thermodynamic quantities, with the enthalpy, the entropy, and the heat capacity displaying very high mutual correlations ($r \gtrsim 0.96$) that follow from standard thermodynamic relations, and moderate correlations with the heat of formation of $r \approx 0.64$, $r \approx 0.62$, and $r \approx 0.57$ respectively. The defect concentration correlates strongly both with the dynamical DCF descriptors ($r \approx 0.78$) and with the heat of formation ($r \approx 0.82$), reinforcing the view that adding vacancies simultaneously raises the energetic cost and enhances the effective dynamical freedom captured by the descriptor. In contrast, the inter-vacancy distance shows comparatively weak linear coupling, with $r \approx 0.18$ against the diffusivity and $r \approx 0.46$ against the heat of formation. The coexistence of a highly redundant block of dynamical descriptors, a tightly coupled thermodynamic block, and a weakly correlated distance variable shows that the relevant information is both redundant and non-linearly encoded, so that a linear model would be limited by multicollinearity. A non-linear model is therefore required, and its capacity must in turn be controlled to avoid overfitting, which motivates the optimization described next.

\begin{figure}[htbp]
\centering
\includegraphics[width=\linewidth]{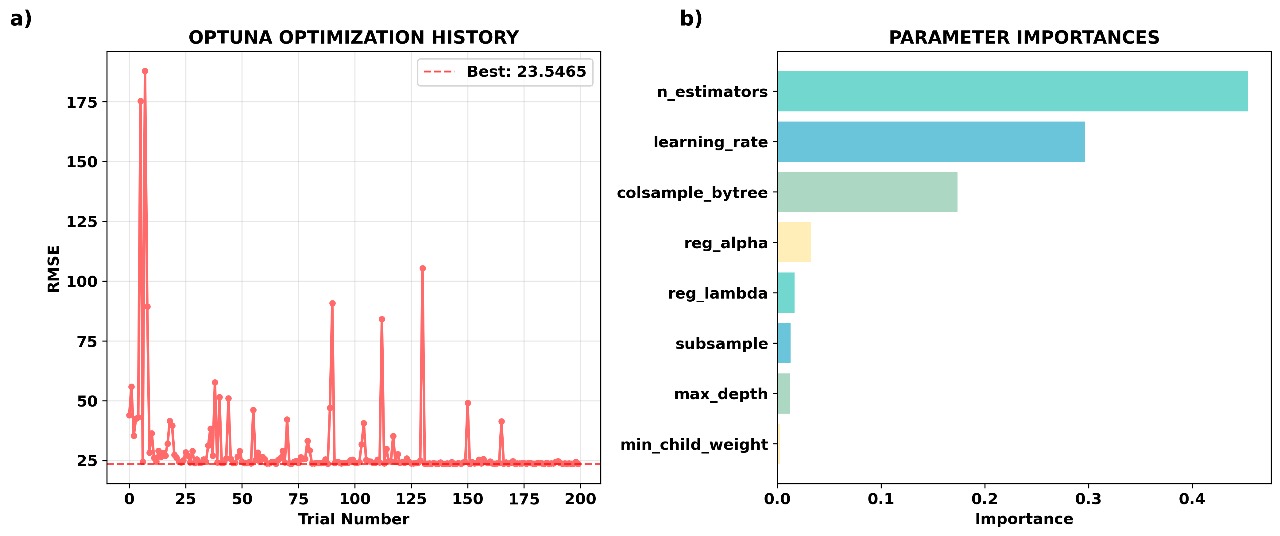}
\caption{Bayesian hyperparameter optimization of the XGBoost model with Optuna. (a) Evolution of the root mean squared error over 500 trials. (b) Relative importance of the optimized hyperparameters.}
\label{fig:optuna_xgb}
\end{figure}

The non-linearity and redundancy exposed by the correlation analysis make the predictive accuracy of the gradient boosted model contingent on a careful calibration of its capacity, which we addressed with Bayesian optimization. Figure~\ref{fig:optuna_xgb} reports the optimization history and the resulting hyperparameter importance.

The optimization history in Figure~\ref{fig:optuna_xgb}(a) follows the root mean squared error over 500 trials. The strong fluctuations of the early trials reflect the exploratory phase of the Bayesian search, and the error converges toward a stable minimum of approximately 23.5~kcal/mol as the search concentrates on the most promising regions of the hyperparameter space. This convergence indicates that the space was sampled efficiently and that the final model corresponds to a well-defined optimum. The hyperparameter importance in Figure~\ref{fig:optuna_xgb}(b) identifies the number of estimators as the most influential parameter, which highlights the role of the ensemble size in controlling the bias--variance trade-off, followed by the learning rate, which governs the stability of the boosting updates. The column-sampling ratio contributes moderately, while the regularization terms, the subsampling ratio, the maximum tree depth, and the minimum child weight contribute less, indicating that the optimal model benefits more from global ensemble control than from strict constraints on individual trees. The optimized hyperparameters were then fixed and used to train the final model employed in the remainder of this work.

\begin{figure}[htbp]
\centering
\includegraphics[width=\linewidth]{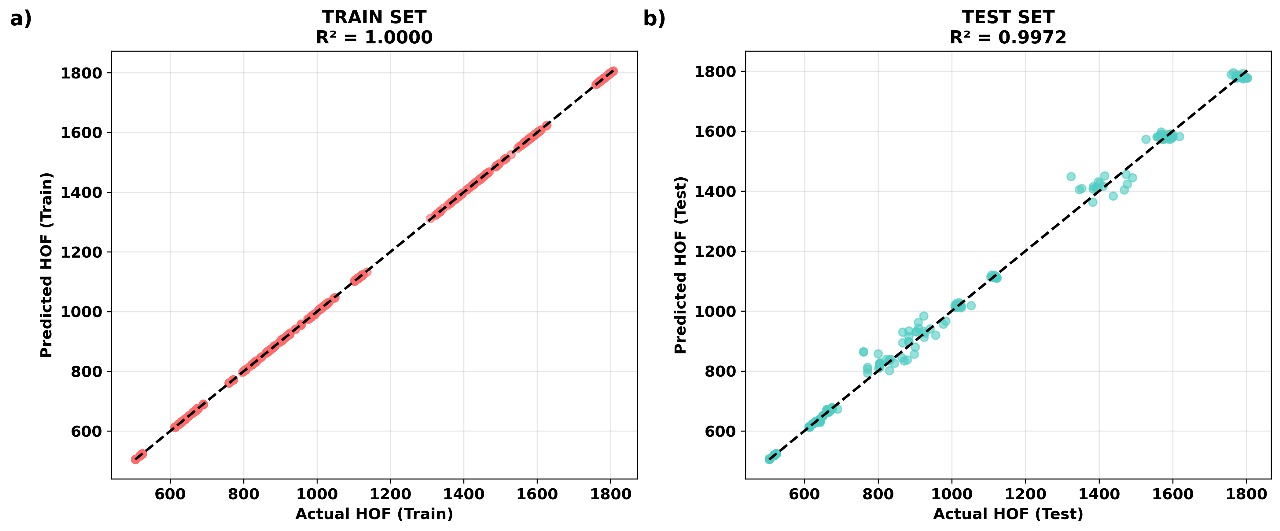}
\caption{Predicted versus calculated heat of formation for the optimized XGBoost model. (a) Training set, 80\% of the data. (b) Independent test set, 20\% of the data.}
\label{fig:xgb_performance}
\end{figure}

With the hyperparameters fixed, the central question is whether the optimized model generalizes beyond the configurations used for training, since only a model that transfers to unseen structures can support mechanistic conclusions. Figure~\ref{fig:xgb_performance} compares the predicted and the calculated heat of formation for the training and the test sets.

The training set in Figure~\ref{fig:xgb_performance}(a) shows the predicted values closely distributed along the ideal diagonal, with a narrow dispersion that confirms that the model captures the dominant non-linear relationships between the descriptors and the stability. The independent test set in Figure~\ref{fig:xgb_performance}(b), which was held out during training, exhibits a comparable dispersion with no systematic bias, and its root mean squared error of approximately 23.5~kcal/mol matches the value reached during optimization, which provides quantitative evidence that the model does not overfit and generalizes to new configurations. The close agreement between the training and the test performance confirms that the combination of a physically informed descriptor, gradient boosted trees, and Bayesian hyperparameter tuning yields a stable and transferable predictor. From a physical standpoint, the simultaneous accuracy on both sets indicates that the heat of formation of defective graphene can be inferred robustly from a small number of structural and dynamical descriptors, which validates the premise that the stability landscape arising from multiple concentrations and separations can be efficiently learned and generalized.

\begin{figure}[htbp]
\centering
\includegraphics[width=\linewidth]{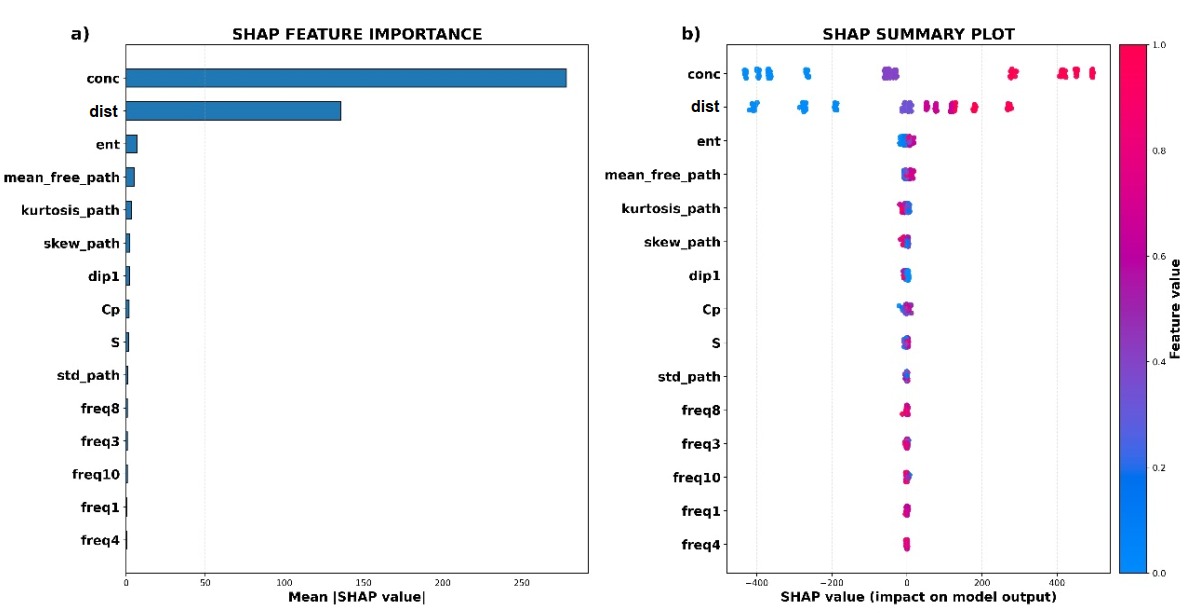}
\caption{SHAP analysis of the optimized XGBoost model. (a) Global importance of the input variables based on the mean absolute SHAP values. (b) SHAP summary plot showing the signed impact of each variable on the predicted heat of formation.}
\label{fig:shap_analysis}
\end{figure}

A model that predicts accurately is not yet an explanation, and the next step is to attribute the prediction to specific physical variables in order to expose the mechanisms that control stability. Figure~\ref{fig:shap_analysis} reports the SHAP analysis of the optimized model, which decomposes each prediction into the signed contributions of the individual variables.

The global importance ranking in Figure~\ref{fig:shap_analysis}(a), based on the mean absolute SHAP values, identifies the defect concentration as the dominant variable by a wide margin, which confirms and extends to the full non-linear response the trend already suggested by the correlation analysis. This dominance reflects the fact that the number of vacancies directly controls the energetic penalty associated with bond breaking, lattice reconstruction, and electronic redistribution. The inter-vacancy distance appears as the second most influential variable, which shows that the spatial arrangement of the vacancies, and not only their number, governs the stability, and that vacancy interactions are not additive with respect to concentration. The remaining features, including the enthalpy, the mean free path, higher-order statistical moments of the DCF path distribution, dipole-related terms, the heat capacity, the entropy, and selected vibrational frequencies, contribute substantially less, but their non-zero contributions confirm that the stability landscape is intrinsically multidimensional and cannot be reduced to one or two variables without loss of information. The SHAP summary plot in Figure~\ref{fig:shap_analysis}(b) resolves these contributions at the level of individual configurations. High concentrations are systematically associated with positive SHAP values and thus with a clear destabilization, while low concentrations shift the prediction toward lower heats of formation, in a monotonic and physically transparent trend. The inter-vacancy distance contributes in the same direction, with larger separations associated with higher heats of formation, a trend whose physical origin is the coalescence of nearby vacancies. When the vacancies are close, they merge and reconstruct into pentagon and heptagon rings that eliminate dangling bonds and lower the energy, whereas at larger separations the defects behave as nearly independent monovacancies that retain their under-coordinated sites and therefore raise the heat of formation \cite{Lee2005, ElBarbary2003, banhart2011structural}. The broad spread of the distance contribution reflects the configuration-dependent nature of this coupling. The remaining features display narrow distributions centered near zero, contributing mainly to the fine tuning of the stability within a given class of concentration and distance. The analysis therefore establishes a clear hierarchy in which the heat of formation is controlled primarily by the defect concentration, secondarily by the inter-vacancy distance, and only at a finer level by the geometric, vibrational, and statistical descriptors, and it confirms that the predictive performance of the model rests on physically meaningful attributes rather than on spurious correlations.

The descriptor hierarchy revealed by the SHAP analysis suggests that the heat of formation should be expressible, to leading order, in terms of the two dominant variables, and we tested this hypothesis by searching for an explicit analytical law. Guided by the dominance of the defect concentration and the inter-vacancy distance, symbolic regression based on genetic programming was applied to distill the learned mapping into a compact expression, which is given by
\begin{equation}
\begin{aligned}
\mathrm{HOF} = {} & (\mathrm{conc} - 1.06)\; e^{\,\mathrm{dist} + \sin(\mathrm{dist} + 0.15) + 2.25} \\
& + 460.20 ,
\end{aligned}
\label{eq:symbolic}
\end{equation}
\noindent where $\mathrm{conc}$ is the defect concentration and $\mathrm{dist}$ is the inter-vacancy distance, and the numerical constants are those returned by the symbolic-regression search, rounded to two decimal places. Equation~\ref{eq:symbolic} captures in closed form the two dominant trends identified above. The concentration enters through a linear prefactor, so that the heat of formation grows essentially in proportion to the number of vacancies, while the distance enters through an exponential term, so that the heat of formation increases with separation in agreement with the coalescence mechanism, with the sinusoidal term accommodating the residual variation among the discrete neighbor shells rather than encoding a genuine periodicity.

\begin{figure}[htbp]
\centering
\includegraphics[width=\linewidth]{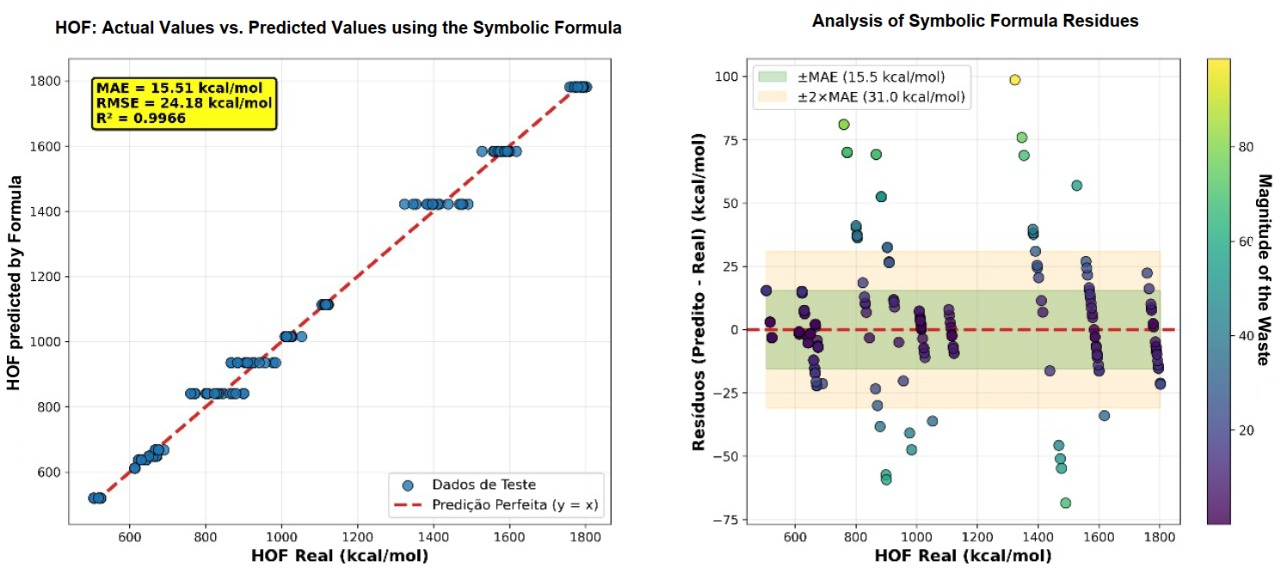}
\caption{Validation of the symbolic-regression model for the heat of formation. (a) Predicted versus reference heat of formation from Equation~\ref{eq:symbolic}. (b) Residuals as a function of the reference heat of formation.}
\label{fig:symbolic_validation}
\end{figure}

The validation of the symbolic model against the independent test set is shown in Figure~\ref{fig:symbolic_validation}. The predicted and the reference heats of formation cluster tightly along the ideal diagonal, with a mean absolute error of 15.51~kcal/mol, a root mean squared error of 24.18~kcal/mol, and a coefficient of determination of $R^{2} = 0.9966$, so that the analytical expression retains essentially the same predictive power as the full gradient boosted model. The residuals lie predominantly within the $\pm\mathrm{MAE}$ and $\pm2\,\mathrm{MAE}$ intervals and show no systematic trend across the full range of the heat of formation, which indicates that the expression provides a uniform description from the weakly to the strongly defective regime. The result establishes that the high-dimensional stability landscape induced by atomic vacancies, first learned by a non-linear ensemble operating on DCF descriptors, can ultimately be condensed into a closed-form law depending on two intuitive variables, which represents a convergence between atomistic simulation, machine learning, and explicit physical modeling.

\section{Conclusions}

We have mapped and rationalized the thermodynamic stability of graphene with atomic vacancies across a broad configurational space. Several hundred defective supercells with systematically varied vacancy concentration and inter-vacancy distance were relaxed with the PM7 Hamiltonian in MOPAC, encoded with the Dynamic Collision Fingerprint, and used to train a gradient boosted decision tree model that reproduces the heat of formation of an independent test set without evidence of overfitting.

The SHAP analysis established a clear hierarchy of mechanisms, with the defect concentration as the dominant factor governing stability, the inter-vacancy distance as the second factor, and the geometric, vibrational, and thermodynamic descriptors providing finer corrections. Notably, the distance dependence is consistent with the coalescence of nearby vacancies into reconstructed pentagon and heptagon rings, which eliminate dangling bonds and stabilize closely spaced defects relative to the isolated monovacancies recovered at larger separations. Symbolic regression compressed this high-dimensional mapping into a compact analytical law that reproduces the heat of formation to within a few percent using only the concentration and the separation, demonstrating that the essential physics of vacancy-induced destabilization can be captured by a low-dimensional yet strongly non-linear relation.

Beyond the specific case of graphene, the workflow establishes a transferable route to defect engineering in two-dimensional materials, in which atomistic modeling, interpretable machine learning, and symbolic regression jointly yield structure--stability relationships that are predictive, physically transparent, and mathematically explicit. This combination paves the way for the rational, data-driven design of defective low-dimensional materials, where stability can be predicted and controlled at a computational cost far below that of direct quantum calculations.


\begin{acknowledgement}
M.L.P.J.\ acknowledges financial support from FAPDF (grant 00193-00001807/2023-16), CNPq (grants 444921/2024-9 and 308222/2025-3), and CAPES (grant 88887.005164/2024-00). Computational resources were provided by the NanoEngineering Laboratory (NanoEng) at the University of Bras\'ilia.

\end{acknowledgement}

\begin{suppinfo}
The data that support the findings of this study are available from the corresponding author upon reasonable request.

\end{suppinfo}

\section*{Author Contributions}
\noindent
M.V.N.C.: Software, Validation, Formal Analysis, Data Curation, and Writing Original Draft. J.R.M.L.F.: Investigation, Software, Validation, and Writing Original Draft. K.R.S.O.: Investigation, Validation, and Writing Original Draft. R.M.T.: Methodology, Software, Validation, and Writing (Review and Editing). M.L.P.J.: Conceptualization, Methodology, Formal Analysis, Resources, Data Curation, Visualization, Writing (Review and Editing), Supervision, Project Administration, and Funding Acquisition.


\section*{Notes}
The authors declare that they have no known competing financial interests or personal relationships that could have appeared to influence the work reported in this paper.


\bibliography{bibliography}

@article{trevethan2014vacancy,
  author  = {Trevethan, Thomas and Latham, Christopher D. and Heggie, Malcolm I. and Briddon, Patrick R. and Rayson, Mark J.},
  title   = {Vacancy diffusion and coalescence in graphene directed by defect strain fields},
  journal = {Nanoscale},
  year    = {2014},
  volume  = {6},
  number  = {5},
  pages   = {2978--2986},
  doi     = {10.1039/C3NR06222H}
}

@article{xiong2021structural,
  author  = {Xiong, Zixin and Zhong, Lei and Wang, Haotian and Li, Xiaoyan},
  title   = {Structural Defects, Mechanical Behaviors, and Properties of Two-Dimensional Materials},
  journal = {Materials},
  year    = {2021},
  volume  = {14},
  number  = {5},
  pages   = {1192},
  doi     = {10.3390/ma14051192}
}

@article{bhatt2022various,
  author  = {Bhatt, Mahesh Datt and Kim, Heeju and Kim, Gunn},
  title   = {Various defects in graphene: a review},
  journal = {RSC Advances},
  year    = {2022},
  volume  = {12},
  number  = {33},
  pages   = {21520--21547},
  doi     = {10.1039/D2RA01436J}
}

@article{gao2019thermodynamic,
  author  = {Gao, Fei and Gao, Shiwu},
  title   = {Thermodynamic stability of magnetic states of monovacancy in graphene revealed by ab initio molecular dynamics simulations},
  journal = {Scientific Reports},
  year    = {2019},
  volume  = {9},
  number  = {1},
  pages   = {751},
  doi     = {10.1038/s41598-018-37333-9}
}

@article{Zhou2013,
  author  = {Zhou, Wu and Zou, Xiaolong and Najmaei, Sina and Liu, Zheng and Shi, Yumeng and Kong, Jing and Lou, Jun and Ajayan, Pulickel M. and Yakobson, Boris I. and Idrobo, Juan-Carlos},
  title   = {Intrinsic Structural Defects in Monolayer Molybdenum Disulfide},
  journal = {Nano Letters},
  year    = {2013},
  volume  = {13},
  number  = {6},
  pages   = {2615--2622},
  doi     = {10.1021/nl4007479}
}

@article{Zhou2017,
  author  = {Zhou, Si and Wang, Shanshan and Li, Huashan and Xu, Wenshuo and Gong, Chuncheng and Grossman, Jeffrey C. and Warner, Jamie H.},
  title   = {Atomic Structure and Dynamics of Defects in 2D MoS2 Bilayers},
  journal = {ACS Omega},
  year    = {2017},
  volume  = {2},
  number  = {7},
  pages   = {3315--3324},
  doi     = {10.1021/acsomega.7b00734}
}

@article{banhart2011structural,
  author  = {Banhart, Florian and Kotakoski, Jani and Krasheninnikov, Arkady V.},
  title   = {Structural defects in graphene},
  journal = {ACS Nano},
  year    = {2011},
  volume  = {5},
  number  = {1},
  pages   = {26--41},
  doi     = {10.1021/nn102598m}
}

@article{lin2015defect,
  author  = {Lin, Zhong and Carvalho, Bruno R. and Kahn, Ethan and Lv, Ruitao and Rao, Rahul and Terrones, Humberto and Pimenta, Marcos A. and Terrones, Mauricio},
  title   = {Defect engineering of two-dimensional transition metal dichalcogenides},
  journal = {2D Materials},
  year    = {2016},
  volume  = {3},
  number  = {2},
  pages   = {022002},
  doi     = {10.1088/2053-1583/3/2/022002}
}

@article{tournus2005vacancy,
  author  = {Tournus, F. and Charlier, J. C.},
  title   = {Vacancy and divacancy in graphene: A first-principles study},
  journal = {Physical Review B},
  year    = {2005},
  volume  = {71},
  number  = {16},
  pages   = {165421},
  doi     = {10.1103/PhysRevB.71.165421}
}

@article{li2021vacancy,
  author  = {Li, Daozhi and Ma, Xiaoyang and Chu, Hongwei and Li, Ying and Zhao, Shengzhi and Li, Dechun},
  title   = {Vacancy-Induced Magnetism in Fluorographene: The Effect of Midgap State},
  journal = {Molecules},
  year    = {2021},
  volume  = {26},
  number  = {21},
  pages   = {6666},
  doi     = {10.3390/molecules26216666}
}

@article{vieirasegundo2016grafeno,
  author  = {Vieira Segundo, J. E. D. and Vilar, E. O.},
  title   = {Grafeno: Uma revis{\~a}o sobre propriedades, mecanismos de produ{\c{c}}{\~a}o e potenciais aplica{\c{c}}{\~o}es em sistemas energ{\'e}ticos},
  journal = {Revista Eletr{\^o}nica de Materiais e Processos},
  year    = {2016},
  volume  = {11},
  number  = {2},
  pages   = {54--57},
  issn    = {1809-8797}
}

@article{vandermeer2017diffusion,
  author  = {van der Meer, Berend and Dijkstra, Marjolein and Filion, Laura},
  title   = {Diffusion and interactions of point defects in hard-sphere crystals},
  journal = {The Journal of Chemical Physics},
  year    = {2017},
  volume  = {146},
  number  = {24},
  pages   = {244905},
  doi     = {10.1063/1.4990416}
}

@article{landesman1985ordering,
  author  = {Landesman, J. P. and Tr{\'e}glia, G. and Turchi, P. and Ducastelle, F.},
  title   = {Electronic structure and pairwise interactions in substoichiometric transition metal carbides and nitrides},
  journal = {Journal de Physique},
  year    = {1985},
  volume  = {46},
  number  = {6},
  pages   = {1001--1017},
  doi     = {10.1051/jphys:019850046060100100}
}

@article{Pietrucci2008,
  author  = {Pietrucci, F. and Bernasconi, M. and Laio, A. and Parrinello, M.},
  title   = {Vacancy-vacancy interaction and oxygen diffusion in stabilized cubic {ZrO2} from first principles},
  journal = {Physical Review B},
  year    = {2008},
  volume  = {78},
  number  = {9},
  pages   = {094301},
  doi     = {10.1103/PhysRevB.78.094301}
}

@article{Bogicevic2003,
  author  = {Bogicevic, A. and Wolverton, C.},
  title   = {Nature and strength of defect interactions in cubic stabilized zirconia},
  journal = {Physical Review B},
  year    = {2003},
  volume  = {67},
  number  = {2},
  pages   = {024106},
  doi     = {10.1103/PhysRevB.67.024106}
}

@article{stewart2013optimization,
  author  = {Stewart, James J. P.},
  title   = {Optimization of parameters for semiempirical methods VI: More modifications to the NDDO approximations and re-optimization of parameters},
  journal = {Journal of Molecular Modeling},
  year    = {2013},
  volume  = {19},
  number  = {1},
  pages   = {1--32},
  doi     = {10.1007/s00894-012-1667-x}
}

@article{stewart2007optimization,
  author  = {Stewart, James J. P.},
  title   = {Optimization of parameters for semiempirical methods V: Modification of NDDO approximations and application to 70 elements},
  journal = {Journal of Molecular Modeling},
  year    = {2007},
  volume  = {13},
  number  = {12},
  pages   = {1173--1213},
  doi     = {10.1007/s00894-007-0233-4}
}

@article{cui2014density,
  author  = {Cui, Qiang and Elstner, Marcus},
  title   = {Density functional tight binding: values of semi-empirical methods in an ab initio era},
  journal = {Physical Chemistry Chemical Physics},
  year    = {2014},
  volume  = {16},
  number  = {28},
  pages   = {14368--14377},
  doi     = {10.1039/c4cp00908h}
}

@article{rapacioli2011extensions,
  author  = {Rapacioli, Mathias and Simon, Aude and Dontot, L{\'e}o and Spiegelman, Fernand},
  title   = {Extensions of DFTB to investigate molecular complexes and clusters},
  journal = {physica status solidi (b)},
  year    = {2012},
  volume  = {249},
  number  = {2},
  pages   = {245--258},
  doi     = {10.1002/pssb.201100615}
}

@article{grimme2019exploration,
  author  = {Grimme, Stefan},
  title   = {Exploration of Chemical Compound, Conformer, and Reaction Space with Meta-Dynamics Simulations Based on Tight-Binding Quantum Chemical Calculations},
  journal = {Journal of Chemical Theory and Computation},
  year    = {2019},
  volume  = {15},
  number  = {5},
  pages   = {2847--2862},
  doi     = {10.1021/acs.jctc.9b00143}
}

@article{pracht2020automated,
  author  = {Pracht, Philipp and Bohle, Fabian and Grimme, Stefan},
  title   = {Automated exploration of the low-energy chemical space with fast quantum chemical methods},
  journal = {Physical Chemistry Chemical Physics},
  year    = {2020},
  volume  = {22},
  number  = {13},
  pages   = {7169--7192},
  doi     = {10.1039/c9cp06869d}
}

@inproceedings{galvao2023confiabilidade,
  author    = {Galv{\~a}o, Breno R. L. and de Carvalho Junior, Nelson Ribeiro},
  title     = {Confiabilidade dos m{\'e}todos semiemp{\'i}ricos e DFTB para o cen{\'a}rio global de otimiza{\c{c}}{\~a}o das estruturas de nanocluster},
  booktitle = {Anais do XII Congresso de Engenharias da Universidade Federal de S{\~a}o Jo{\~a}o del-Rei (COEN)},
  year      = {2023},
  doi       = {10.29327/1461105.12-6}
}

@article{butler2018machine,
  author  = {Butler, Keith T. and Davies, Daniel W. and Cartwright, Hugh and Isayev, Olexandr and Walsh, Aron},
  title   = {Machine learning for molecular and materials science},
  journal = {Nature},
  year    = {2018},
  volume  = {559},
  number  = {7715},
  pages   = {547--555},
  doi     = {10.1038/s41586-018-0337-2}
}

@article{schmidt2019recent,
  author  = {Schmidt, Jonathan and Marques, M{\'a}rio R. G. and Botti, Silvana and Marques, Miguel A. L.},
  title   = {Recent advances and applications of machine learning in solid-state materials science},
  journal = {npj Computational Materials},
  year    = {2019},
  volume  = {5},
  number  = {1},
  pages   = {83},
  doi     = {10.1038/s41524-019-0221-0}
}

@article{birschitzky2022machine,
  author  = {Birschitzky, Viktor C. and Ellinger, Florian and Diebold, Ulrike and Reticcioli, Michele and Franchini, Cesare},
  title   = {Machine learning for exploring small polaron configurational space},
  journal = {npj Computational Materials},
  year    = {2022},
  volume  = {8},
  number  = {1},
  pages   = {125},
  doi     = {10.1038/s41524-022-00805-8}
}

@article{ong2021accelerating,
  author  = {Ong, Shyue Ping},
  title   = {Accelerating materials science with high-throughput computations and machine learning},
  journal = {Computational Materials Science},
  year    = {2019},
  volume  = {161},
  pages   = {143--150},
  doi     = {10.1016/j.commatsci.2019.01.013}
}

@article{Tromer2025,
  author  = {Tromer, Raphael M.},
  title   = {Dynamic Collision Fingerprints (DCF): Introducing a New Descriptor Linking Lattice Interactions to 2D Structural Data Signatures},
  journal = {Journal of Chemical Theory and Computation},
  year    = {2025},
  volume  = {21},
  number  = {16},
  pages   = {8106--8118},
  doi     = {10.1021/acs.jctc.5c00856}
}

@article{Tromer2026,
  author        = {Tromer, R. and Felix, I. M. and Besse, R. and Junior, M. L. P. and Luz, M. G. E.},
  title         = {A Comparative Study of Structural Representations for 2D Materials: Insights from Dynamic Collision Fingerprint and Matminer},
  journal       = {arXiv preprint},
  year          = {2026},
  eprint        = {2602.22950},
  archivePrefix = {arXiv}
}

@article{Choudhary2018,
  author  = {Choudhary, Kamal and DeCost, Brian and Tavazza, Francesca},
  title   = {Machine learning with force-field inspired descriptors for materials: Fast screening and mapping energy landscape},
  journal = {Physical Review Materials},
  year    = {2018},
  volume  = {2},
  number  = {8},
  pages   = {083801},
  doi     = {10.1103/PhysRevMaterials.2.083801}
}

@article{Dau2023,
  author  = {Dau, Minh Tuan and Al Khalfioui, M. and Michon, A. and Reserbat-Plantey, Antoine and V{\'e}zian, S.},
  title   = {Descriptor engineering in machine learning regression of electronic structure properties for 2D materials},
  journal = {Scientific Reports},
  year    = {2023},
  volume  = {13},
  number  = {1},
  pages   = {5426},
  doi     = {10.1038/s41598-023-31928-7}
}

@article{NaTalang2026,
  author  = {Na Talang, Cheewawut and Kesorn, A. and Cholsuk, Chanaprom and Vogl, Tobias and Hunkao, Rutchapon},
  title   = {Chemical Feature Engineering and Defect-Aware Structural Fingerprint Representations for Complex Defects in 2D Materials},
  journal = {Journal of Chemical Information and Modeling},
  year    = {2026},
  volume  = {66},
  number  = {4},
  pages   = {2017--2029},
  doi     = {10.1021/acs.jcim.5c02100}
}

@inproceedings{chen2016xgboost,
  author    = {Chen, Tianqi and Guestrin, Carlos},
  title     = {{XGBoost}: A Scalable Tree Boosting System},
  booktitle = {Proceedings of the 22nd ACM SIGKDD International Conference on Knowledge Discovery and Data Mining},
  year      = {2016},
  pages     = {785--794},
  doi       = {10.1145/2939672.2939785}
}

@inproceedings{lundberg2017unified,
  author    = {Lundberg, Scott M. and Lee, Su-In},
  title     = {A Unified Approach to Interpreting Model Predictions},
  booktitle = {Advances in Neural Information Processing Systems (NeurIPS)},
  year      = {2017},
  volume    = {30},
  pages     = {4765--4774}
}

@article{lundberg2020local,
  author  = {Lundberg, Scott M. and Erion, Gabriel and Chen, Hugh and DeGrave, Alex and Prutkin, Jordan M. and Nair, Bala and Katz, Ronit and Himmelfarb, Jonathan and Bansal, Nisha and Lee, Su-In},
  title   = {From Local Explanations to Global Understanding with Explainable AI for Trees},
  journal = {Nature Machine Intelligence},
  year    = {2020},
  volume  = {2},
  number  = {1},
  pages   = {56--67},
  doi     = {10.1038/s42256-019-0138-9}
}

@article{akiba2019optuna,
  author    = {Akiba, Takuya and Sano, Shotaro and Yanase, Toshihiko and Ohta, Takeru and Koyama, Masanori},
  title     = {{Optuna}: A Next-generation Hyperparameter Optimization Framework},
  journal   = {Proceedings of the 25th ACM SIGKDD International Conference on Knowledge Discovery and Data Mining},
  year      = {2019},
  pages     = {2623--2631},
  doi       = {10.1145/3292500.3330701}
}

@article{schmidt2009distilling,
  author  = {Schmidt, Michael and Lipson, Hod},
  title   = {Distilling Free-Form Natural Laws from Experimental Data},
  journal = {Science},
  year    = {2009},
  volume  = {324},
  number  = {5923},
  pages   = {81--85},
  doi     = {10.1126/science.1165893}
}

@article{cranmer2020discovering,
  author  = {Cranmer, Miles and Sanchez-Gonzalez, Alvaro and Battaglia, Peter and Xu, Rui and Cranmer, Kyle and Spergel, David and Ho, Shirley},
  title   = {Discovering Symbolic Models from Deep Learning with Inductive Biases},
  journal = {arXiv preprint arXiv:2006.11287},
  year    = {2020}
}

@article{guimera2020scientific,
  author  = {Guimer{\`a}, Roger and Reichardt, Ignasi and Aguilar-Mogas, Antoni and Massucci, Francesco A. and Miranda, Manuel and Pallar{\`e}s, Jordi and Sales-Pardo, Marta},
  title   = {A Bayesian machine scientist to aid in the solution of challenging scientific problems},
  journal = {Science Advances},
  year    = {2020},
  volume  = {6},
  number  = {5},
  pages   = {eaav6971},
  doi     = {10.1126/sciadv.aav6971}
}

@article{brunton2016discovering,
  author  = {Brunton, Steven L. and Proctor, Joshua L. and Kutz, J. Nathan},
  title   = {Discovering Governing Equations from Data by Sparse Identification of Nonlinear Dynamical Systems},
  journal = {Proceedings of the National Academy of Sciences},
  year    = {2016},
  volume  = {113},
  number  = {15},
  pages   = {3932--3937},
  doi     = {10.1073/pnas.1517384113}
}

@article{udrescu2020ai,
  author  = {Udrescu, Silviu-Marian and Tegmark, Max},
  title   = {{AI Feynman}: A Physics-Inspired Method for Symbolic Regression},
  journal = {Science Advances},
  year    = {2020},
  volume  = {6},
  number  = {16},
  pages   = {eaay2631},
  doi     = {10.1126/sciadv.aay2631}
}

@article{ElBarbary2003,
  author  = {El-Barbary, A. A. and Telling, R. H. and Ewels, C. P. and Heggie, M. I. and Briddon, P. R.},
  title   = {Structure and energetics of the vacancy in graphite},
  journal = {Physical Review B},
  year    = {2003},
  volume  = {68},
  number  = {14},
  pages   = {144107},
  doi     = {10.1103/PhysRevB.68.144107}
}

@article{Lee2005,
  author  = {Lee, G.-D. and Wang, C. Z. and Yoon, E. and Hwang, N.-M. and Kim, D.-Y. and Ho, K. M.},
  title   = {Diffusion, Coalescence, and Reconstruction of Vacancy Defects in Graphene Layers},
  journal = {Physical Review Letters},
  year    = {2005},
  volume  = {95},
  number  = {20},
  pages   = {205501},
  doi     = {10.1103/PhysRevLett.95.205501}
}

@article{Thomas2022,
  author  = {Thomas, D. M. and Asiri, Y. and Drummond, N. D.},
  title   = {Point defect formation energies in graphene from diffusion quantum Monte Carlo and density functional theory},
  journal = {Physical Review B},
  year    = {2022},
  volume  = {105},
  number  = {18},
  pages   = {184114},
  doi     = {10.1103/PhysRevB.105.184114}
}

@article{Yazyev2007,
  author  = {Yazyev, O. V. and Helm, L.},
  title   = {Defect-induced magnetism in graphene},
  journal = {Physical Review B},
  year    = {2007},
  volume  = {75},
  number  = {12},
  pages   = {125408},
  doi     = {10.1103/PhysRevB.75.125408}
}

@article{Krasheninnikov2010,
  author  = {Krasheninnikov, A. V. and Nordlund, K.},
  title   = {Ion and electron irradiation-induced effects in nanostructured materials},
  journal = {Journal of Applied Physics},
  year    = {2010},
  volume  = {107},
  number  = {7},
  pages   = {071301},
  doi     = {10.1063/1.3318261}
}

@article{Bartok2013,
  author  = {Bart{\'o}k, Albert P. and Kondor, Risi and Cs{\'a}nyi, G{\'a}bor},
  title   = {On representing chemical environments},
  journal = {Physical Review B},
  year    = {2013},
  volume  = {87},
  number  = {18},
  pages   = {184115},
  doi     = {10.1103/PhysRevB.87.184115}
}

@article{Behler2007,
  author  = {Behler, J{\"o}rg and Parrinello, Michele},
  title   = {Generalized Neural-Network Representation of High-Dimensional Potential-Energy Surfaces},
  journal = {Physical Review Letters},
  year    = {2007},
  volume  = {98},
  number  = {14},
  pages   = {146401},
  doi     = {10.1103/PhysRevLett.98.146401}
}

@article{Deringer2017,
  author  = {Deringer, Volker L. and Cs{\'a}nyi, G{\'a}bor},
  title   = {Machine learning based interatomic potential for amorphous carbon},
  journal = {Physical Review B},
  year    = {2017},
  volume  = {95},
  number  = {9},
  pages   = {094203},
  doi     = {10.1103/PhysRevB.95.094203}
}

\end{document}